\documentclass[preprint,groupedaddress]{revtex4}

\usepackage{graphicx}
\usepackage{epstopdf}
\usepackage{color}
\usepackage{amsfonts,amssymb}
\usepackage{natbib}
\usepackage[applemac]{inputenc}

\begin{document}

\title{Acceleration statistics of material particles in turbulent flow}

\author{Nauman M. Qureshi, Unai Arrieta, Christophe Baudet, Alain Cartellier, Yves Gagne, Micka\"el Bourgoin}

\address{Laboratoire des \'Ecoulements G\'eophysiques et Industriels, CNRS/UJF/INPG UMR5519, BP53, 38041 Grenoble, France}

\begin{abstract}
Being able to accurately model and predict the dynamics of dispersed inclusions transported by a turbulent flow, remains a challenge with important scientific, environmental and economical issues. One critical and difficult point is to correctly describe the dynamics of particles over a wide range of sizes and densities. Our measurements show that acceleration statistics of particles dispersed in a turbulent flow do exhibit specific size and density effects but that they preserve an extremely robust turbulent signature with lognormal fluctuations, regardless of particles size and density. This has important consequences in terms of modelling of the turbulent transport of dispersed inclusions.
\pacs{47.27.Gs, 47.27.T-, 82.70.-y}
\end{abstract}

\maketitle

Turbulent transport of material inclusions plays an important role in many natural and industrial processes. In marine ecosystem for instance, interactions between the turbulent sea and the gametes of marine animals is determinant for the efficiency of the reproduction and the spreading of the species \cite{bib:denny2002_BiolBull}. Industrial stakes concern mixing and combustion, but also pollutants dispersion. Being able to accurately model and predict the dynamics of dispersed inclusions transported by a turbulent flow, remains a challenge with important scientific, environmental and economical issues. One critical and difficult point is to correctly describe particles' dynamics over a wide range of sizes and densities. 
The dynamics of particles transported by a turbulent flow is known to be affected by size and density effects. If particles are neutrally buoyant and small (namely smaller than the smallest turbulent eddies, at the energy dissipation scale $\eta$ of the flow), they behave as fluid tracers and their dynamics reflects fluid particles dynamics. This property is commonly used to characterize single phase flows from tracer particles imaging techniques (PIV, LDV, PTV). However, when particles density is larger or smaller than the surrounding fluid and/or when particles size become comparable to turbulent eddies, their dynamics deviates from that of fluid particles and tends to be affected by so called \emph{inertial effects}. Among them some can be qualitatively described in terms of particles interaction with turbulent eddies. For instance, centrifugal forces are generally expected to cluster heavier particles in low vorticity regions (and lighter in high vorticity regions), an effect known as \emph{preferential concentration}. However, from a quantitative point of view no reliable model has emerged yet to describe and predict accurately the statistical properties of particles advected by a turbulent flow. For instance we are unable today to predict correctly the dynamics of water droplets in a cloud \cite{bib:shaw2003}, what considerably limits our understanding of rain mechanisms. Actually, even writing an appropriate equation of motion for a particle transported by a turbulent flow remains a theoretical challenge which has only been approached in some limit cases, assuming generally point like particles \cite{bib:gatignol1983,bib:maxey1983,bib:bec2006_JFM,bib:balkovsky2001_PRL,bib:Zaichik2004}; but the range of validity of such models for real particles, with finite size and finite density, as well as the minimal relevant ingredients for a model to be pertinent remains unclear.

\begin{figure}[b!]
\includegraphics[width=8cm]{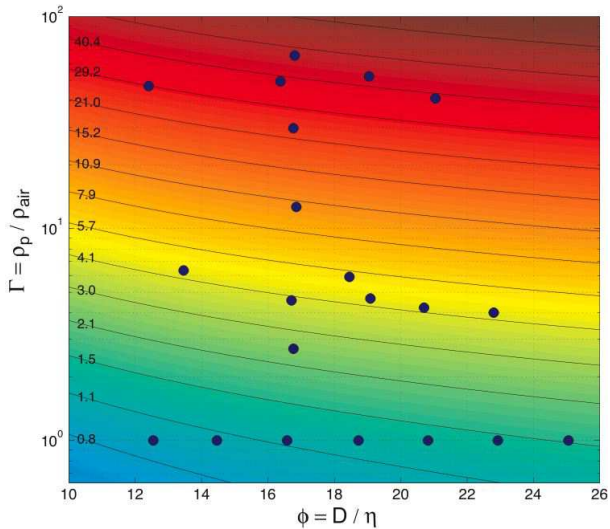}
\caption{Particles classes considered in the present study, described in the $(\phi,\Gamma)$ phase space. 
The countour lines indicate an estimation of the particles Stokes number $St=\tau_p/\tau_D$, where the respone time $\tau_p$ of the particle has been corrected for added mass and for finite Reynolds number effects according to Schiller and Nauman \cite{bib:clift1978} and $\tau_D$ is the flow eddy turnover time at the scale of the particle diameter $D$.}\label{fig:PhiGamma}
\end{figure}

In the present study we have measured acceleration statistics of isolated material particles transported in a turbulent air flow, varying systematically particles size and density. Acceleration is a kinematic quantity of particular interest since it directly reflects the forces exerted by the carrier flow on the particles, which is the basic ingredient of any dynamical model for the particles equation of  motion. As particles we use soap bubbles ; their density can be adjusted from neutrally buoyant to about 70 times heavier than air. As their Weber number is extremely small, they are known to not deform and to behave as rigid spheres. The seeding density is extremely low (particles are injected individually) so that particles can be considered as isolated and do not backreact on the carrier flow. Particles are therefore only characterized by two parameters : the ratio $\phi$ of their diameter $D$ to the dissipative scale of the flow ($\phi = D/\eta$) and the ratio $\Gamma$ of their density $\rho_{\textrm{\small p}}$ to the carrier fluid (air in our case) density $\rho_{\textrm{\small air}}$ ($\Gamma = \rho_{\textrm{\small p}}/\rho_{\textrm{\small air}}$). Figure~\ref{fig:PhiGamma} summarizes all the particle classes (in the $(\phi,\Gamma)$ parameter space) that we have considered in the present  study. As far as we know this represents the most exhaustive exploration over such a wide range of particles sizes and densities ever done in a same experimental configuration.

Our experiment runs in a large wind tunnel with a measurement section of $0.75~\textrm{m} \times 0.75~\textrm{m}$ where the turbulence is generated downstream a grid with a mesh size of 6~cm and reproduces almost ideal isotropic turbulence. The results reported here were obtained with a mean velocity of the fluid $U=15\;\textrm{m}\cdot\rm{s}^{-1}$ and a turbulence level $u_{rms}/U \simeq 3\%$. The corresponding Reynolds number, based on Taylor microscale, is of the order of $R_\lambda=160$. The dissipation scale $\eta=(\nu^3/\epsilon)^{1/4}$ is 240~$\mu$m (where $\nu$ is the kinematic viscosity of air and $\epsilon$ the turbulent energy dissipation rate per unit mass) and the energy injection scale $L$ is 6~cm. Particles are individually tracked using 1D Lagrangian acoustic Doppler velocimetry \cite{bib:gervais2007_ExpFluids,bib:mordant2005_RSI,bib:mordant2002_JASA}. We measure the streamwise velocity component $v_z$ of the particles as they are tracked along their trajectory. The streamwiase acceleration component $a_z$ is obtained by differentiating the velocity using a convolution with a differentiated gaussian kernel \cite{bib:mordant2004_PhysicaD}. Each particle is tracked during approximately 50 ms, which corresponds to several dissipation time scales $\tau_{\eta}=(\nu/\epsilon)^{1/2} \simeq 3.8 ms$. For each point in the $(\phi,\Gamma)$ parameter space we record at least 4000 tracks at a samplimg rate of 32768~Hz, giving more than $10^6$ data points per set.

\begin{figure}[t!]
\includegraphics[width=9.5cm]{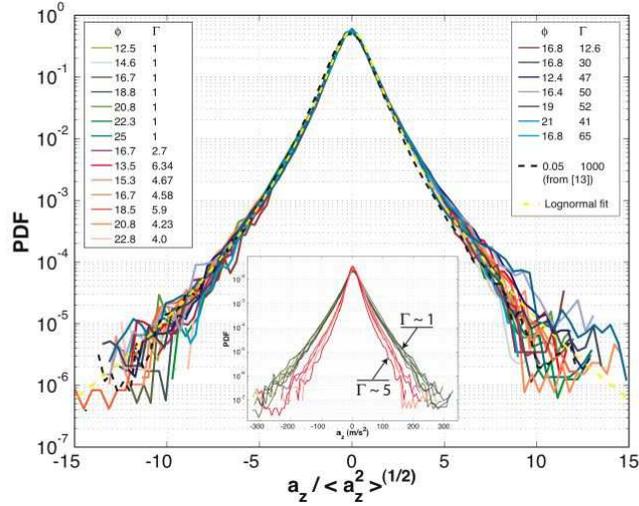}
\caption{Normalized component acceleration probability density function (PDF) of material particles transported in a turbulent air flow. Dashed line corresponds to the measurement by Ayyasomayalula et al. \cite{bib:ayya2006} for water droplets (with size $\phi\sim5\times10^{-2}$ and $\Gamma\sim1000$). Dot-dashed line is a fit by the relation ${\cal {P}}(x)=\frac{e^{3s^2/2}}{4\sqrt{3}} \left[ 1- {\rm erf} \left(\frac{ {\rm ln}\left(\left| x / \sqrt{3} \right| \right)+2s^2} {\sqrt{2}s} \right) \right]$ associated to a lognormal distribuion of the acceleration amplitude \cite{bib:mordant2004b,bib:qureshi2007_PRL} (best fit is found for $s\sim0.62$, corresponding to a distribution flatness ${\cal{F}}=\frac{9}{5}e^{4s^2}\sim8.4$). Inset shows the non normalized component acceleration PDF for $\Gamma\sim1$ (outer bundle, green tone) and $\Gamma\sim5$ (inner bundle, red tone) particles and various values of $\phi$.
}\label{fig:AccPDFn}
\end{figure}

Figure \ref{fig:AccPDFn} represents the probability density functions (PDF), of the acceleration component normalized to variance one, ($a_z/\left<a_z^2\right>^{1/2}$), for all studied particles. Remarkably we find that, within statistical errorbars in the rare events tails, all the PDFs almost collapse onto a single curve, indicating that normalized acceleration statistics depends significantly neither on particle size, nor on particle density. These results extend to the case of heavy particles our previous observation for neutrally buoyant particles \cite{bib:qureshi2007_PRL}, where we have shown this PDF to be correctly described by a robust lognormal distribution $\cal P$ (yellow dot dashed line on figure \ref{fig:AccPDFn}, see the caption for its analytical expression). They are also consistent with recent measurements, in a water von K\'arm\'an flow, for particles in the range $0.5<\Gamma\lesssim2$ \cite{bib:volk2007_EPL,bib:volk2008_PhysicaD}.

Although for technical reasons, we cannot reduce at present the bubble's diameter below $\phi \sim 10$, it is still interesting to  compare our results with other recent measurements obtained by Z. Warhaft's group at Cornell University for heavy sub-kolmogorov particles ($\phi\ll1$)\cite{bib:ayya2006}. Using high speed optical tracking, they have measured acceleration statistics of small water droplets ($\phi \sim 5\times 10^{-2}$ ; $\Gamma \sim 1000$ ; $St \sim 0.1$ ) transported in a turbulent air flow with comparable characteristics to ours (both are windtunnel experiments with similar isotropy levels), although at a slightly higher Reynolds numbers ($R_\lambda \sim 250$). We have superimposed on figure \ref{fig:AccPDFn} the normalized acceleration PDF from their measurements (black dashed line). We find it to be almost undistinguishable from our measurements with larger particles. This suggests that acceleration is a physical quantity with an extremely robust lognormal statistical signature, over a very wide range of sizes ($D \sim 0.05\eta \rightarrow L/10$) and densities ($\Gamma \sim 0.5 \rightarrow 1000$). Once normalized to variance one, all PDFs are well fitted by the same distribution $\cal P$ which is only parametrized by its flatness $\cal F$ (in the present experiments ${\cal{F}}\simeq8.4$). 

A deeper insight into the specific effects associated with particles' size and density can be obtained by examining acceleration variance. Since the statistics of acceleration normalized to variance one is found essentially independent of particles size and density, an effect of those parameters can indeed only be expected to affect the acceleration variance itself $\left<a_z^2\right>$. The inset in figure \ref{fig:AccPDFn} represents the true acceleration PDF (not normalized to variance one) for neutrally buoyant ($\Gamma\sim1)$ and heavy ($\Gamma\sim 5$) particles, with varying sizes. It appears that heavy particles have a much narrower and peaked PDF, indicating a gobal decrease of acceleration variance with increasing density. The spreading of the curves for a given density indicates that size also influences particles acceleration variance.

\begin{figure}[t!]

\includegraphics[width=9cm]{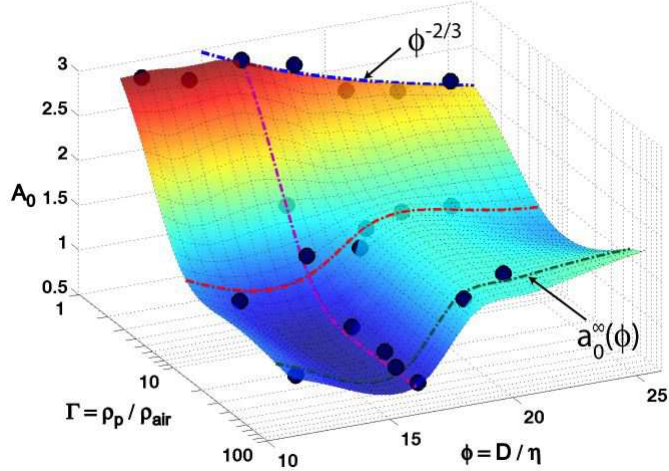}
\caption{Acceleration variance for all particles studied in the $(\phi,\Gamma)$ parameter space (dots). The surface represents a rough interpolation based on the available measurements. The dot dashed lines materialize the interpolation along the best resolved constant $\phi$-lines and $\Gamma$-lines. For the $\Gamma=1$ set, the dot-dashed line coincide with a $\phi^{-2/3}$ decay for $\phi\gtrsim 15$ \cite{bib:qureshi2007_PRL}.}\label{fig:A0_surf}
\end{figure}

\begin{figure}[t!]
\includegraphics[width=8.5cm]{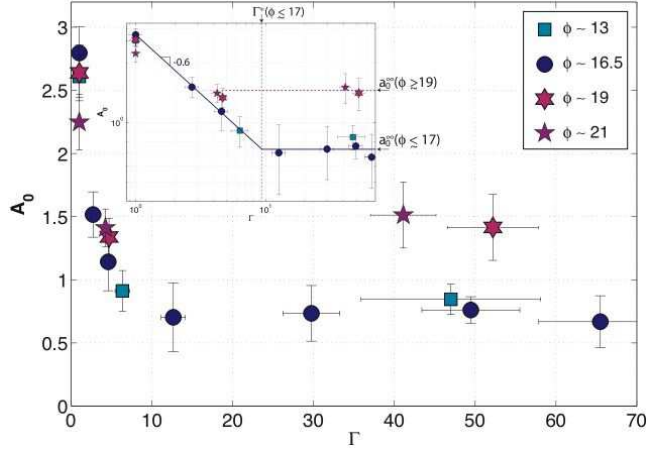}
\caption{2D-Projection  on the $(\Gamma-A_0)$ plane  of measurements data points on figure \ref{fig:A0_surf}, showing the evolution of acceleration variance with density ratio $\Gamma$ for different particle sizes ($\phi\sim16.5$ set (circles) is the best resolved). The inset shows the same data in a log-log plot. Errorbars are mostly due to experimental uncertainties in the determination of particles size and density.}\label{fig:Arms_Gamma}
\end{figure}

A closer analysis of acceleration variance shows a non-trivial dependence with size and density. We report in figure \ref{fig:A0_surf} the acceleration variance for all the measurements we have performed (in the following, we consider the dimensionless acceleration variance normalized according to Heisenberg-Yaglom's scaling : $A_0(\phi,\Gamma)=\left<a_z^2\right>\epsilon^{-3/2}\nu^{1/2}$). The mapping of the $(\phi,\Gamma)$ parameter space that we have been able to achieve, allows us to infer a rough interpolation of the evolution of $A_0(\phi,\Gamma)$, represented by the surface in figure \ref{fig:A0_surf}. Though only the coarse tendency of this interpolation is relevant, several important qualitative features can still be observed. For small and close to neutral particles $A_0 \sim 2.8$, a value consistent with previous measurements \cite{bib:voth2002} and DNS \cite{bib:vedula1999_PoF,bib:gotoh2001_PRL} for particles in the fluid tracer limit. If we consider the effect of an increasing density at a fixed particle size (along $\phi=\textrm{constant-lines}$), $A_0$ is always found to decrease and to saturate to a finite limit (noted $a_0^\infty(\phi)$) for the largest densities. This is better quantified in a 2D projection (figure \ref{fig:Arms_Gamma}). The $\phi \sim 16.5$ set of measurements (which is our most complete set of density effects at fixed size) shows that the evolution of acceleration variance with density exhibits two different regimes (see inset in figure  \ref{fig:Arms_Gamma}): (i) for low densities $A_0(\phi,\Gamma)=a_0(\phi)\Gamma^\alpha$ ($a_0(\phi)$ corresponds to size effects for the neutrally buoyant case which we have previously studied in \cite{bib:qureshi2007_PRL}) with $\alpha\sim0.6$ (ii) for large densities $A_0$ saturates to the finite limit $a_0^\infty(\phi)$ ($a_0^\infty(16.5)\sim0.7$). The transition between these two regimes occurs for a characteristic density ratio $\Gamma^*(\phi)$ ($\Gamma^*(16.5)\sim10$). As shown on figure \ref{fig:Arms_Gamma} from the measurements at different values of $\phi$, the existence of these two regimes seems to hold for all studied particle sizes but with a size dependent transition density ratio $\Gamma^*(\phi)$ and a size dependent saturation value $a_0^\infty(\phi)$. 
If we consider more closely size effects at fixed particle density (along $\Gamma=\textrm{constant-lines}$), as seen on figure \ref{fig:A0_surf}, the scenario is actually rather complex, since depending on the density ratio $\Gamma$, $A_0$ can either decrease or increase with particle's size. For neutrally buoyant particles ($\Gamma=1$) $A_0$ starts to deviate from the fluid tracer value for particles larger than about $\phi\sim15$ and then decreases monotically as $\phi^{-2/3}$. We have shown in a previous study \cite{bib:qureshi2007_PRL} that this is the expected scaling for inertial range sized particles when the main forcing simply comes from the spatial pressure differences of the unperturbed flow around the particle. As we move to larger density ratios we then observe a continuous transition toward a drastically changed size dependence for the largest densities ($\Gamma > \Gamma^*$) where ~$a_0^\infty(\phi)$ experience a sudden increase for sizes around $17<\phi<19$. 
Outside this transition region (i.e. for $\phi \lesssim 17$ and $\phi \gtrsim 19$), $a_0^\infty$ doesn't exhibit significant dependence on $\phi$ (at least in the accessible range of sizes) as also seen on figure \ref{fig:Arms_Gamma}, where we have $a_0^\infty(13)\simeq a_0^\infty(16.5)$ and $a_0^\infty(19)\simeq a_0^\infty(21)$. 
Inset in figure \ref{fig:Arms_Gamma} also suggests that $\Gamma^*(\phi)$ might exhibit a similar transition in the same range of sizes as we have $\Gamma^*(13)\simeq\Gamma^*(16.5) \sim 10$ and $(\Gamma^*(19), \Gamma^*(21))< 5$. A size dependence of the exponent $\alpha$ (for $\Gamma<\Gamma^*(\phi)$) cannot be excluded, though a more detailed mapping of the parameter space in the region $\Gamma < \Gamma^*(\phi)$ is still required to be conclusive.

Our measurements have important consequences in terms of the development of accurate models for the turbulent transport of finite size material particles. The robustness of normalized acceleration PDF, regardless of particles' size and density, imposes a strong constraint on the statistical properties of the forcing terms to be included in the equation of motion. For instance, models based on the Maxey \& Riley equation \cite{bib:gatignol1983,bib:maxey1983} for inertial point particles (characterized mainly by their Stokes number) predict a monotonic decrease of acceleration flatness, with a continuous trend of acceleration PDF to gaussianity, as well as a monotonic decrease of acceleration variance with increasing particles Stokes number \cite{bib:bec2006_JFM,bib:volk2008_PhysicaD} (which in these models is only parametrized by an increasing particle's response time and eventually interpreted as an increasing density ratio and/or particle size). These trends are contradictory with the experimental evidence we report here (acceleration flatness remains size and density independent and acceleration variance may increase with particle size in the limit of high density ratios). 
Our results indicate that the high Stokes number extrapolation of such point particle models is not adequate to correctly describe the dynamics of finite size inertial particles. In other words, inertial range sized particles cannot be correctly modeled by simply considering increasing response times in the point-particle approximation; density and finite size effects require a simultaneous specific modeling.

With this goal, finally, we briefly discuss and propose a simple phenomenology for the observed trends of acceleration variance with size and density in the context of the recently introduced \textit{sweep-stick} mechanism \cite{bib:chen2006_JFM,bib:goto2008_PRL} which offers an interesting frame to interpret several of our observations. Physically this mechanism relies on the simple fact that inertial particles reside longer in the quietest regions of the flow and tend therefore to stick preferentially near zero-acceleration points of the carrier turbulent field (this eventually leads to the so-called preferential concentration effect, responsible for particles clustering \cite{bib:squires1991_PoF,bib:aliseda2002}, which in this context is not only attributed to centrifugation by high vorticity regions) along which they tend then to be advected.  This phenomenology is consistent with the decrease of acceleration variance that we observe when particles density increases. Another consistent point is the fact that velocity statistics of our particles (not shown here) are found identical to that of the carrier flow (obtained from classical hotwire anemometry), a feature also shown numerically for zero-acceleration points \cite{bib:goto2008_PRL} (but which is at odds from usual predictions based on point-like particles approaches~\cite{bib:deutsch1991}). Pushed further, the \textit{sweep-stick} phenomenology also offers a consistent frame to possibly explain the sudden increase of acceleration variance with particle size observed for high density ratios: while heavy and small enough particles can indeed be expected to efficiently \textit{hide} in the quietest regions of the flow, when particles become larger than the typical size $L^*$ of these regions, the quietening effect is damped as particles experience again the influence of active regions in the turbulent field. Though the existence of such a typical scale is still controversial, recent measurements of preferential concentration suggest that the quiet sticking regions might have a characteristic size $L^*$ in the range $10-20\eta$ \cite{bib:aliseda2002} consistent with the $17\lesssim\phi\lesssim19$ range for which we observe the sudden increase of $a_0^\infty(\phi)$.

To summarize, using a versatile material particle generator we have been able to explore the simultaneous influence of size and density ratio on the turbulent transport of material particles. 
Their acceleration statistics are found to be robustly described by a lognormal distribution, where only the variance depends significantly on particle's size and density. We have shown that, for heavy particles, finite size effects can be trivially extrapolated neither from the heavy point particle case nor from the finite size neutrally buoyant case. However, the simultaneous influence of density and size may be consistently interpreted in the context of \textit{sweep-stick} mechanisms, what should stimulate further numerical investigations of acceleration field in turbulent flows. It is our hope that these measurements will contribute bridging experimental observations and developments of accurate models for the turbulent transport of particles.\\

\begin{acknowledgments}
\end{acknowledgments}
We deeply acknowledge  E. Calzavarini, J.-F. Pinton, R. Volk, and Z. Warhaft for fruitful discussions.
 
\bibliographystyle{apsrev}
\bibliography{main}
 
\end{document}